\documentclass[12pt]{article}  
\newcommand{\comma}{\;\; , \; \; }
\newcommand{\period}{\;\; .}
\newcommand{\eq}{\; = \;}
\newcommand{\sep}{\;\; , \;\;}

\newcommand{\be}{\begin{equation}}
\newcommand{\bd}{\begin{displaymath}}
\newcommand{\ee}{\end{equation}}
\newcommand{\ed}{\end{displaymath}}
\newcommand{\ba}{\begin{eqnarray}}
\newcommand{\ea}{\end{eqnarray}}
\newcommand{\trace}{{\rm trace} \; }
\newcommand{\Kb}{\overline{K}}
\newcommand{\minus}{\! - \!}

\renewcommand{\i}{{\rm i}}

\newcommand{\e}{{\rm e}}

\renewcommand{\arraystretch}{1.5}

\newcommand{\Q}{{Q}}
\renewcommand{\i}{{\bf  i}}

\title{Algebraic reduction of the Ising model}

\author{ R.J. Baxter\\
{\protect \small  Mathematical
Sciences Institute,  The Australian National}\\
{\protect  \small  University,
 Canberra, A.C.T. 0200, Australia, \small e-mail: none }}

\date{}

\begin{document}


\maketitle

 \abstract{We consider the Ising model on a cylindrical lattice
of $L$ columns, with fixed-spin boundary conditions on the top 
and bottom rows. The spontaneous magnetization can be written
in terms of partition functions on this lattice.
We show how we can use the Clifford algebra of 
Kaufman to write these partition functions in terms of $L$ by $L $
determinants, and then further reduce them to $m$ by $m$ 
determinants, where $m$ is approximately $L/2$.
In this form the results can be compared with those of
the Ising case of the superintegrable chiral Potts model. 
They point to a way of 
calculating the spontaneous magnetization  of that more general 
model  algebraically.}


 \vspace{5mm}




 \section{Introduction}
\setcounter{equation}{0}

Onsager calculated the partition function of the two-dimensional 
square-lattice Ising model in 1944.\cite{Onsager1944} He did so 
by observing that the row-to-row transfer matrices generated a finite 
Lie algebra. This method was refined by Kaufman in 1949
\cite{Kaufman1949}, who showed that the matrices could be expressed 
in terms of a Clifford algebra. By taking the thermodynamic limit of a large 
lattice, one  obtains the free energy per site.

The calculation of the  spontaneous  magnetization is much harder than 
that of the free energy.  Onsager first announced his famous result  for 
the  spontaneous  magnetization at a conference at 
Cornell University in  August 1948 and later at a IUPAP conference in 
Florence.\cite{Onsager1949,MPW1963}  Onsager never published his 
derivation - the puzzle was finally solved by Yang in 1952.\cite{Yang1952}


Here we wish to re-visit this problem, with a view to seeing if the techniques
can be generalized to  the $N$-state superintegrable case of the 
solvable chiral Potts 
model,  which has properties very similar to the Ising model
and reduces to it when $N=2$. The author has previously
used large-lattice functional relations and analytic methods to obtain the 
spontaneous magnetization of the general solvable chiral 
Potts model\cite{Baxter2005a,Baxter2005b}, but it would still be interesting 
to have an algebraic method that may give more insight into the properties 
of the model on a finite lattice.


In section 2 we define the Ising model on a cylindrical lattice
of $L$ columns, with fixed-spin boundary conditions on the top 
and bottom rows. We show how the magnetization
can be expressed, firstly as the ratio of partition functions,
and consequently  as a matrix element between the two ground-state
eigenvectors of the transfer matrix $T$. In section 3 we introduce the
hamiltonian $\cal H$ which commutes with $T$ and define
``hamiltonian partition functions" $\tilde{Z}, \widetilde{W}$. These  are  
limiting cases of the usual partition functions and are slightly simpler 
to work with.  This $\tilde{Z}$ corresponds to the usual partition function,
$\widetilde{W}$ to the partition function with an extra factor $\sigma_1$,
where $\sigma_1$ is some spin deep inside the lattice.

In section 4 we set up the apparatus of the Clifford algebra and use it to 
write $\tilde{Z}, \widetilde{W}$ as square roots of  $L$ by $L$ determinants.
Up to this point our calculation parallels that of Yang.\cite{Yang1952}
Montroll {\it et al}\cite{MPW1963} also calculated the spontaneous 
magnetization using the more combinatorial pfaffian method and 
Szeg{\H o}'s theorem on Toeplitz  matrices.\cite{Szego1958}
In section 5 we marry Yang and Montroll {\it et al}'s techniques by 
using Szeg{\H o}'s theorem to evaluate the appropriate large-$L$  limit 
of $\tilde{W}$ and  hence  derive
Onsager's famous formula for the spontaneous magnetization.

In section 6 we show how the expressions for
$\tilde{Z}, \widetilde{W}$  
can be reduced to linear expressions in determinants of dimension
$[(L- \Q  )/2]$, where $\Q = 0 $ or 1. Then $\tilde{Z}$ is precisely
the hamiltonian limit of the corresponding partition function
of the two-state superintegrable chiral Potts model. In  section
7 we write the result for $\widetilde{W}$  in 
terms of two orthogonal matrices $B_+, B_-^{}$.

In section 8 we comment on whether our result for $\widetilde{W}$ can be 
generalized to the $N$-state superintegrable chiral Potts model.




 \section{The model}
 
 
\setcounter{equation}{0}
 
 The model is defined on the square lattice $\cal L$, rotated through 
$45^{\circ}$, with $M+1$ horizontal rows, each containing  $L$ spins, 
as in Fig. 1.


 \setlength{\unitlength}{1pt}
 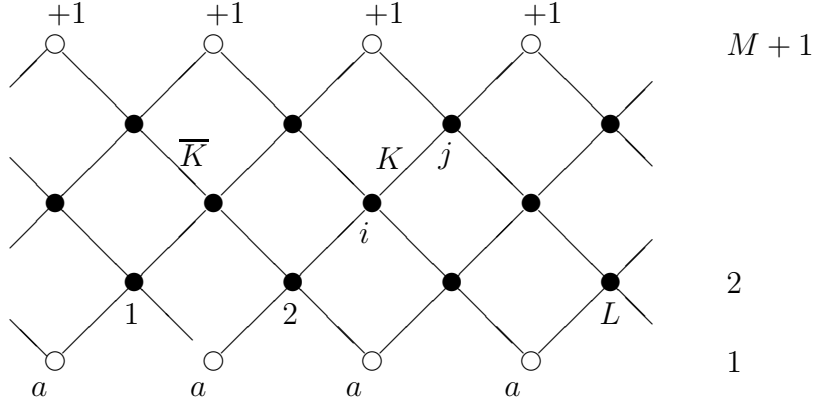
\begin{figure}[hbt]
 \begin{picture}(400,160) (-23,17)

 \put (43,43) {\line(1,1) {74}}
 
 \put (123,63) {\line(1,1) {54}}
 \put (183,63) {\line(1,1) {54}}
 \put (243,63) {\line(1,1) {43}}
 \put (63,3) {\line(1,1) {54}}
 \put (123,3) {\line(1,1) {54}}
 \put (183,3) {\line(1,1) {54}}
 \put (243,3) {\line(1,1) {43}}
 \put (43,104) {\line(1,1) {14}}
 
 \put (63,117) {\line(1,-1) {54}}
 \put (123,117) {\line(1,-1) {54}}
 \put (183,117) {\line(1,-1) {54}}
 \put (243,117) {\line(1,-1) {43}}
   
 \put (43,77) {\line(1,-1) {69}}
 \put (123,57) {\line(1,-1) {54}}
 \put (183,57) {\line(1,-1) {54}}
 \put (243,57) {\line(1,-1) {43}}
  \put (43,16) {\line(1,-1) {14}}

 \multiput(60,0)(60,0){4}{\circle{7}}
 \multiput(60,60)(60,0){4}{\circle*{7}}
 \multiput(60,120)(60,0){4}{\circle{7}}
 \multiput(90,30)(60,0){4}{\circle*{7}}
 \multiput(90,90)(60,0){4}{\circle*{7}}

   \put (51,-13) {$a$}
 \put (111,-13) {$a$}
 \put (170,-13) {$a$}
 \put (230,-13) {$a$}
 
   \put (86,14) {$1$}
   \put (146,14) {$2$}
   \put (266,14) {$L$}

   \put (57,128) {$+1$}
    \put (117,128) {$+1$}
   \put (177,128) {$+1$}
   \put (237,128) {$+1$}

 \put (175,45) {$i$}
 \put (205,75) {$j$}
  \put (181,73) {$K$}
  
  \put (314,-4) {$1$}
  \put (314,26) {$2$}
  \put (314,116) {$M+1$}
 \put (107,74) {$\overline{K}$}
 
 \end{picture}
 \vspace{1.5cm}
 \caption{\footnotesize The square lattice $\cal L$  turned 
 through $45^{\circ}$.}
 
  \label{sqlatt45}
 \end{figure}


 We impose cylindrical boundary 
 conditions, so that the last column $L$ is followed by the first column 
 1. At each site  $i$ there is a spin $\sigma_i$, taking the values
 $+1$ or $ -1$. The spins in the bottom row are fixed to have 
 value $a = \pm 1$, those in the top row to have value $+1$. Adjacent spins  
 $\sigma_i, \sigma_j$ on southwest to northeast edges (with $i$ below 
 $j$) interact with Boltzmann weight $\exp(K \sigma_i  \sigma_j)$;  
 those on southeast to northwest edges with weight 
 $\exp(\Kb  \sigma_i \sigma_j)$.
   
  An important associated parameter is 
 \be \label{defkp}
 k' \eq \left(  \sinh 2K \sinh 2 \Kb  \right) ^{-1}  \period \ee
 The system is ferromagnetically ordered if $K, \Kb$ are positive real and 
 \be 0 < k' < 1 \period \ee
 To align with our notation for the chiral Potts 
 model\cite{Baxter2005a,Baxter2005b} we are defining the RHS of
 (\ref{defkp}) to be $k'$, rather than $k$ as in 
 Onsager.\cite[eq. 2.1a]{Onsager1944}
 
 \subsubsection*{Partition function}
 
   The partition function, which depends on $a$,  is 
   \be Z_a  \eq \sum_{\sigma} \prod_{\langle i,j \rangle} 
\exp (K \sigma_i \sigma_j)  \prod_{\langle i,j \rangle} 
  \exp ( \Kb \sigma_i \sigma_j)   \comma \ee
   the products being over all edges of the two types. The sum is 
  over all values of all the free spins.
    The partition function  can be written as
   \be Z_a \eq u_a^{\dagger} \, { T} ^M \, u_+ \comma \ee
   where ${ T}$ is the row-to-row transfer matrix, with elements
   \be \label{defT}
   T_{\sigma, \sigma'} \eq  \prod_{i=1}^L  \exp (K \sigma_i \sigma'_{i+1})  
  \exp ( \Kb  \sigma_i \sigma'_i)
  \comma  \ee
   $\sigma$ being the set of all spins $\sigma_1, \ldots , \sigma_L$ 
  in one row, and $\sigma'$ being the set in the row above. Thus  
  $ T$ is a $2^L$ by $2^L$ matrix. The vector $u_a$ is of 
  dimension $2^L$, with entries
   \ba ( u_a)_{\sigma}  & = & 1\;\;  {\rm if }  \;\; \sigma_1 = 
   \cdots  = \sigma_L = a \comma \nonumber \\
     & = & 0 \; \; {\rm otherwise} \period \ea

 Two transfer matrices $T, T'$, with different values of $K$ and $ \Kb$, 
 commute  provided they have the same value of 
 $k'$.\cite[section 7.5]{Baxter1982}

Here $u^{\dagger}$ denoted the transposed conjugate of $u$. However,
all our matrices are real, and $T$ commutes with its transpose, so all the 
eigenvectors we shall discuss are also real: the 
complex conjugation is unnecessary.

We shall also need the matrices $S_1, \ldots , S_L$, $C_1, 
\ldots , C_L$, defined by
\be
(S_j)_{\sigma,\sigma'} \eq \sigma_j \prod_{n=1}^L 
\delta(\sigma_n,\sigma'_n) \comma \ee
\be
(C_j)_{\sigma,\sigma'} \eq  \delta(\sigma_j,-\sigma'_j) 
{\prod_{n=1}^L}^{\! \raisebox{-10pt}{*}}  \delta(\sigma_n,\sigma'_n) 
\comma \ee
where the $*$ means that the term $n=j$ is excluded from the
product, 
If  $I, S, C$ are the two-by-two matrices
\renewcommand{\arraystretch}{1.2}
\setlength{\arraycolsep}{4pt}
\be I = \left( \begin{array}
{cc} 1 & 0 \\
0  & 1 \end{array} \right) \sep
S = \left( \begin{array}
{cc} 1 & 0 \\
0  & -1 \end{array} \right) \sep
C = \left( \begin{array}
{cc} 0 & 1 \\ 1 & 0 \end{array} \right) \comma \ee
then we can write $S_j, C_j$ as
\ba S_j  & = &  I \otimes \cdots \otimes I \otimes S \otimes I \otimes \cdots 
\otimes I \nonumber \\ 
C_j  & = & I \otimes \cdots \otimes I \otimes C \otimes I \otimes \cdots 
\otimes I \comma \nonumber \ea
the $S, C$ on the RHS being in position $j$ in the sequence of  $L$ factors.



 \subsubsection*{Spontaneous magnetization}
 

 Take $a= +1$, so all top and bottom boundary spins are $+$.
Let $0$ be a site deep within the lattice. The expectation value of its spin
 $\sigma_0$ is
 \be \label{defM}
{\cal M} \eq \langle \sigma_0 \rangle   \eq Z_+^{-1} \, 
 \sum_{\sigma} \sigma_0  \prod_{\langle i,j \rangle} 
\exp (K \sigma_i \sigma_j)  \prod_{\langle i,j \rangle} 
  \exp ( \Kb \sigma_i \sigma_j) 
\period  \ee
We take the limit when the lattice is 
infinitely large, so
$L, M \rightarrow \infty$, and site $0$ is infinitely far 
from the boundaries.

The ${\langle i,j \rangle} $ products are unchanged by  negating all 
spins $\sigma_i$, so if we imposed toroidal boundary conditions, then it 
would be true that
\be \langle \sigma_0  \rangle =   - \langle \sigma_0 \rangle   \ee
and this would imply that
${\cal M} = 0$. 
At high temperatures ($k' \geq 1$), this is true also for our
fixed-spin boundary conditions when we take the large-lattice
limit. However,  at lower temperaturers  ($0 < k' < 1$)
the system has ferromagnetic long-range order and ``remembers''
the boundary conditions even in the limit of $0$ deep 
inside a large lattice, so that
\be 0 < {\cal M} < 1 \period \ee

If $0$ is in column $i$ of row $j+1$, then we can write 
(\ref{defM}) as
\be \label{defM2}
 {\cal M} \eq  u_+^{\dagger}  T^j S_i T^{M-j} u_+  / 
u_+^{\dagger} T^M u_+ \ee
Because of our cylindrical boundary conditions, we can cyclically 
permute the columns of $\cal L$, so the RHS must be
 independent of  $i$.


   
    \subsubsection*{The sub-spaces}


 The operator that negates all the spins in a row is
   \be \label{defR}
   R \eq  C_1 C_2 \cdots C_L \period   \ee
  We can divide the full $2^L$-dimensional space into two orthogonal 
  subspaces   ${\cal V}_+$,   ${\cal V}_-^{}$ such that
  \be R \,  v =  r v \; \; {\rm if \; \; }  v \in {\cal V}_r  \sep r \eq + \;\; {\rm or} \; \; - 
   \ee
  The operator $R$ commutes with $T$ and $\cal H$, so its eigenvectors 
  lie either in  ${\cal V}_+$ or  ${\cal V}_-^{}$. Let $\Lambda_a$ be the 
  largest eigenvalue of $T$ in ${\cal V}_a$,  $\psi_a$ the corresponding 
  eigenvector normalized so that $\psi_a^{\dagger} \psi_a = 1$,
  and  define
   \be \label{defvpm}
   v_+ \eq (u_+ + u_-)/\sqrt{2} \sep  v_- \eq (u_+ - u_-)/\sqrt{2}  \period \ee
   Then $T^j v_a \in {\cal V}_a$ for all $j$ and
   \be v_a^{\dagger} T^M v_b = 0 \; \; {\rm if \; \; } a \neq b \comma \ee
 \be v_a^{\dagger} T^j S_i T^{M-j} v_b = 0 \; \; {\rm if \; \; } a  = b \comma \ee
 so (\ref{defM2}) becomes
 \be \label{defM3}
  {\cal M} \eq \frac{v_+^{\dagger} T^j S_i T^{M-j} {v}_{- }^{}+ v_-^{\dagger} 
 T^j S_i T^{M-j} v_+}{v_+^{\dagger} T^M  v_+  +  v_-^{\dagger} T^M v_-^{}}
 \period \ee
 
 
   
    \subsubsection*{Asymptotic degeneracy}

If $j, M-j$ are large we can take 
$ T ^j =  \Lambda_+^j  \psi_{+} 
   \psi_{+}^{\dagger} $  or $ \Lambda_-^{j} \psi_{-}^{} 
   \psi_{-}^{\dagger}$ (depending on the sub-space in which $T$ is acting).
   Similarly for $T^{M-j}$ and $T^M$.
  Thus the four matrix elements in (\ref{defM3}) will be 
 proportional to
$\Lambda_+^j \Lambda_-^{M-j}$,  $\Lambda_-^j \Lambda_+^{M-j}$,
$\Lambda_+^M$, $\Lambda_-^M$, respectively. However, for $0 <k <1$, 
$\Lambda_+$ and $\Lambda_-^{}$ are {\it asymptotically degenerate}, 
in that for $L$ large
\be\label{simp}
 \Lambda_+/\Lambda_-^{} = 1 + O(\e^{-L\zeta }) \sep 
(v_+^{\dagger} \psi_+ ^{} ) /(v_-^{\dagger} \psi_- ^{} )   = 1 + O(\e^{-L\zeta }) 
 \comma \ee
choosing the signs of the $\psi_a$ appropriately. Here $\zeta$ is
 independent of $L$ and is a measure of the interfacial tension
between the two phases.\cite[section 7.10]{Baxter1982}

That this should be so can be seen from a low-temperature series
expansion of the eigenvalues $\Lambda$ and eigenvectors $\psi$
of $T$ (or more conveniently the $\cal H$ of the next sub-section) 
 in increasing powers of $k'$. 
 
One can start from a single
zero-temperature configuration in which either all spins are $+1$, or
all are $-1$ ($u_+$ or $u_-^{}$). The expansion then proceeds
identically for each choice (going from one choice to the other by 
negating all spins) until one reaches terms of order $k'^L$, when one 
has to consider the opposite state (all spins $-1$ or all $+1$) from the 
original. This has the same eigenvalue for $k'=0$, so naive 
eigenvalue perturbation theory fails. Then and only then does one  
have to  decide whether to symmetrize or anti-symmetrize the 
eigenvector with respect to $R$.

 If  $\phi_+, \phi_-^{}$ 
are the corresponding two near-eigenvectors, then up to this order
the eigenvalues are the same and 
\be u_-^{\dagger} \phi_+^{}  =   u_+^{\dagger} \phi_-^{} = 0 \period \ee
The actual eigenvectors of $T$ are obtained by symmetrizing:
\be \psi_+ = (\phi_+ + \phi_-^{})/\sqrt{2}  \sep \psi_- = 
(\phi_+ - \phi_-^{})/\sqrt{2}   \period \ee
The relations (\ref{simp}) then follow, using (\ref{defvpm}).

  
Hence in the limit of $j, M-j, L$ all large we can take 
   $ \Lambda_+ = \Lambda_-^{}$, 
     and (\ref{defM3})
  becomes
  \be \label{defM4}
    {\cal M} \eq   \psi_+^{\dagger} S_i \psi_-^{}  \period \ee



 \section{The  associated hamiltonian $H$}
 
 
\setcounter{equation}{0}



Consider the limit when $K $ becomes small and $\Kb$ becomes large 
while $k'$ remains fixed. Then to first order in $K$ we obtain 
from (\ref{defT})
\be \e^{-L \Kb} \, T    =  {\cal I}  - K {\cal H}  = \e^{-K {\cal H}}  \comma \ee
where $\cal I$ is the $2^L$-dimensional identity matrix,
\be \label{HHH}
{ \cal H} \eq   { \cal H} _0 +k' { \cal H} _1 \comma \ee
where
\be \label{defH01}
{\cal H}_0  =  - \sum_{i=1}^{L}   S_i S_{i+1} 
\sep {\cal H}_1  =  - \sum_{i=1}^{L}  C_i 
 \ee
and $S_{L+1} = S_1$. It follows that this hamiltonian $\cal H$ commutes 
with the transfer matrix
$T$ and with $R$. Its ground state eigenvector (the one corresponding 
to the most negative eigenvalue of $\cal H$ , and to the largest eigenvalue 
of $T$) in the
subspace ${\cal V}_a$ will therefore be $\psi_a$.

When $k' = 0$ , the hamiltonian reduces to the diagonal matrix
${\cal H}_0$ and we see that the ground state  eigenvectors  
are indeed $u_+$ and  $u_-^{}$, both with eigenvalue $-L$. it is 
convenient to define the closely related
 matrix
 \be \label{defJc}
 {\cal J} =  {\cal H}_0 + L {\cal I} = \sum_{i=1}^{L}  ({\cal I} - S_i S_{i+1} ) 
 \period  \ee
 This has minimum eigenvalue $0$, the corresponding eigenvectors being
 $v_+$ and $v_-^{}$.
 

   
 \subsubsection*{Hamiltonian partition functions}


We could continue to look at the four matrix elements in (\ref{defM3}), 
or attempt to evaluate directly the result (\ref{defM4}). We prefer to follow an 
intermediate path and to consider the expressions
\be \label{defZ}
\tilde{Z}_{+} (\alpha) \eq v_{+}^{\dagger} \e^{-\alpha {\cal H}}
 v_{+} \sep  \tilde{Z}_{-}^{} (\beta) \eq v_{-}^{\dagger} \e^{-\beta {\cal H}}
 v_{-}^{}  \comma \ee
 \be  \label{defW}  
 \widetilde{W} (\alpha,\beta, x) \eq  v_+^{\dagger} \e^{-\alpha {\cal H}} 
 \e^{-\rho {\cal J}} S_1
 \e^{-\beta {\cal H}}  v_-^{} \comma  \ee
 where
 \be x = \e^{-4 \rho} \period \ee
 This corresponds to taking the limits $K \rightarrow 0$,
 $j, M-j \rightarrow \infty$ in the matrix elements of (\ref{defM3}), 
 while keeping  $\alpha = jK$ and $\beta = K(M-j)$ fixed.

  The advantage of these expressions is that they are finite-lattice 
 partition functions, but with $T^j$ replaced by the rather simpler
 matrix $\e^{-\alpha {\cal H}}$.   For definiteness, we have taken
  the $i $ in (\ref{defM3}) to be 1.

We have introduced the factor  $ \e^{-\rho {\cal J}} $ into $\widetilde{W}$ 
because the elements of the diagonal matrix $\cal J$ take the values 
$0,4, 8, \ldots , 4 [L/2]$, where $[x]$ is the integer part of $x$.
Hence $ \widetilde{W} (\alpha,\beta, x)$ is a polynomial 
in $x$  of degree $[L/2]$. This naturally manifests itself in the following
working and provides a useful check against errors. 

 If $\rho \rightarrow \infty$, so $x \rightarrow 0$, then
 \be  \e^{-\rho {\cal J}} S_1 \rightarrow 
 u_+ u_+^{\dagger} - u_-^{} u_-^{\dagger}  = 
  v_+ v_-^{\dagger} +v_-^{} v_+^{\dagger} \comma \ee
 hence
  \be \widetilde{W} (\alpha,\beta, 0) \eq   \tilde{Z}_{+} (\alpha) 
  \,  \tilde{Z}_{-} ^{} (\beta) \period \ee
  Also,
  \bd  \widetilde{W} (\alpha,0, x) \eq   \tilde{Z}_{+} (\alpha) 
\comma \ed
 \be  \widetilde{W} (0,\beta, x) \eq   \tilde{Z}_{-}^{} (\beta) 
\period \ee


When $\alpha$ is large and positive, we can replace 
$\e^{-\alpha {\cal H}}$ in the above definitions by
$\e^{-\alpha \Lambda_+} \psi_+ \psi_+^{\dagger}$ or by
$\e^{-\alpha \Lambda_-^{}} \psi_-^{} \psi_-^{\dagger}$,
according to the sub-space in which $\cal H$ is acting. One can then verify 
that 
\be  \label{resM}
 {\cal M} \eq \lim_{\alpha, \beta, L \rightarrow \infty} \; 
\frac { \widetilde{W} (\alpha,\beta, 1) }
{[ \tilde{Z}_{+} (2 \alpha)  \, \tilde{Z}_{-}^{} (2 \beta) ]^{1/2}} \period \ee

 We show in the following sections that we can use the techniques of 
 Kaufman\cite{Kaufman1949} and Yang\cite{Yang1952}  to
 express the $\tilde{Z}, \widetilde{W}$ expressions as square roots of 
 $L$ by $L$  determinants, and hence evaluate $\cal M$ by using 
 Szeg{\H o}'s theorem on Toeplitz forms.\cite{Szego1958,MPW1963}
  We further 
 show  that the determinants themselves can be expressed as squares 
 of  $m$ by $m$ determinants, where
 $L-2 \leq 2 m \leq L$.




 \section{The Clifford algebra}
 
 
\setcounter{equation}{0}




\subsubsection*{General remarks}

Set \be D_j \eq    \i \, C_j S_j \eq - \i S_j C_j \comma  \ee
\be \label{defG}
\Gamma_j = C_1 C_2 \cdots C_{j-1} D_j \sep 
\Gamma_{L+j} = C_1 C_2 \cdots C_{j-1} S_j \ee
for $j = 1, \ldots, L$.

These  $\Gamma_1, \ldots ,\Gamma_{2L}$ anti-commute:
\be  \Gamma_i  \Gamma_j +  \Gamma_j  \Gamma_i = 
2 \, \delta_{i,j} \,  {\cal I} \ee
for $1 \leq i,j \leq 2L$, so form a Clifford algebra. If $P$ is any orthogonal
(not necessarily real) $2 L$ by $2 L$ matrix and
\be \label{trgr}
\Gamma^*_i \eq \sum_{j=1}^{2L} P_{i,j} \Gamma_j \comma \ee
then it is also true that 
\be  \Gamma^*_i  \Gamma^*_j +  \Gamma^*_j  \Gamma^*_i = 
2 \, \delta_{i,j} \,  {\cal I}  \ee

If $V$ is a $2^L$-dimensional matrix such that
\be \label{group}
V \Gamma_i V^{-1} \eq \sum_{j=1}^{2L} v_{j,i} \Gamma_j \comma \ee
for $i = 1, \ldots 2L$, then  we say that the $2L$ by $2L$ 
matrix $\widehat{V}$ with
elements $v_{i,j}$ is the {\it representative} of $V$. If $V$ is real and 
symmetric, then $\widehat{V}$ is orthogonal and hermitian.

Such matrices form a group ${\cal G}$.\cite{Baxter1977} If $V, V'$ have the 
property (\ref{group}), with representatives $\widehat{V}, \widehat{V'}$, 
 then so does  $V V'$, and its representative is $\widehat{V} \widehat{V'}$.  
 Also, if  $\widehat{V}= \widehat{V'}$, then
\bd V = c \, V' \comma \ed
where $c$ is a scalar factor. This means that a representative
matrix determines its parent to within a scalar factor.

If $\Gamma_i, \Gamma^*_i$ are related by  (\ref{trgr}), then
\be V \Gamma^*_i V^{-1} = \sum_{j} v^*_{j,i} \Gamma^*_j \comma 
\ee
where the $v^*_{i,j} $ are the elements of the
matrix $ P \widehat{V} P^{T} $. Thus mapping  $\Gamma_i $ to
$\Gamma^*_i$ takes $\widehat{V}$ to $ P \widehat{V} P^{T} $.
Similary, if a  matrix $2^L$-dimensional $H$ satisfies
\be  H \Gamma_i - \Gamma_i H =  (H, \Gamma_i ) \eq 
\sum_{j=1}^{2L} h_{j,i} \Gamma_j   \comma \ee
where the $h_{i,j}$ are elements of a matrix $\widehat{H}$,  then the
 same  mapping takes  $\widehat{H}$ to   $ P \widehat{H} P^{T} $. 
 We call $\hat{H}$ the {\it H-representative} of $H$.

Any representative
 $\widehat{V}$ is orthogonal and any $H$-representative $\widehat{H}$ is 
anti-symmetric. If $V$ ($H$) is hermitian, then (because the 
$\Gamma_j$ are hermitian), then
$\widehat{V}$ ($\widehat{H}$) is also hermitian.



\subsubsection*{Exponentials of $\cal H$, $\widehat{\cal H}$}


If $\widehat{H}$ is hermitian, then it must be pure imaginary and 
anti-symmetric.
 It is diagonalizable by a unitary
matrix. Further, its eigenvalues are real and
occur in pairs $\lambda$ and $ -\lambda$, 
the corresponding eigenvectors being complex conjugates of one another.
Adding and subtracting these eigenvectors, it follows that there is a real
orthogonal transformation that takes  $\widehat{\cal H}$ to the form
\be {\cal D} \eq  \left( \begin{array}
{cc} 0 &  \i D  \\ -\i D  & 0 \end{array} \right) \comma \ee
where $D$ is a  diagonal matrix.

By making such a transformation and then explicitly considering
the two-by-two constituent blocks of $\cal D$, we find that
$\exp(-\alpha {\cal H}) \in \cal G$, so  
\be  \label{reprH}
\; \; {\rm   representative \; of } \;  
\exp(-\alpha {\cal H}) = 
 \exp( - \alpha \widehat{\cal H} ) \period \ee

\subsubsection*{Partition functions}
 If $V$ is hermitian, we can similarly find an orthogonal
 transformation (\ref{trgr}) that takes $\widehat{V}$ to block diagonal form, 
 each block being hermitian and orthogonal, so of the form
 \be   \label{repV}
 \left( \begin{array}
{cc} \cosh 2 \alpha_j  & \i \sinh 2 \alpha_j \\ - \i \sinh 2 \alpha_j
 & \cosh 2 \alpha_j \end{array} \right)   \ee
 where $j = 1, \ldots , L$.
 If we arrange the $\Gamma^*_j$ so that this block is in rows and columns
 $j, L+j$, it follows that
 \be V\eq c  \, \exp \left( \i \sum_{j=1}^L \alpha_j \Gamma^*_j  
  \Gamma^*_{L+j} \right)  \comma \ee
  where $c$ is some scalar factor.
  
  Each term $\Gamma^*_j  
  \Gamma^*_{L+j}$ commutes with all the other terms, so we have 
  decomposed $V$ into a direct product of two-by-two matrices, 
  the $j$th such matrix having eigenvalues
   $\e^{\alpha_j}, \e^{-\alpha_j}$. Hence
   \be
\trace V / ( \det V )^{1/d} \eq \prod_{j=1}^L 2 \, \cosh \alpha_j 
\comma \ee
where $d = 2^L$. On the other hand, from (\ref{repV}),
\be
\det  \left( I + \widehat{V}  \right) \eq  \prod_{J=1}^L 4 \cosh^2 \alpha_j 
\comma \ee
so we have the identity
\be \label{trace}
 \label{trV} \trace V / ( \det V )^{1/d}
\eq \left[ \det ( I + \widehat{V}) \right]^{1/2}
\period \ee
Note that the relations (\ref{reprH}), (\ref{trV}) are quite general, being
unchanged by the orthogonality transformation (\ref{trgr}).



\subsubsection*{Representative of $\exp(-\alpha {\cal H} )$ }


{From}  (\ref{defH01}), (\ref{defG}),
\be \label{quad}
{\cal H}_0 \eq \i R\,  \Gamma_L \Gamma_{L+1}  - \i \sum_{j=1}^{L-1} 
\Gamma_j \Gamma_{L+j+1} 
\comma \ee
\bd {\cal H}_1 \eq  \i \sum_{j=1}^{L} \Gamma_j \Gamma_{L+j} 
\comma \ed
so we see that $ {\cal H}_0 $, $ {\cal H}_1$ are both quadratic forms
in the $\Gamma_i$, provided we introduce the spin-reversal 
operator $R$. We can do this, since in either sub-space ${\cal V}_{\pm}$,
$R$ commutes with $T$ and $\cal H$, having the value $r  = +1$ in 
${\cal V}_{+}$, and $r = -1$ in ${\cal V}_{-}^{}$.

{From} now on we take  ${\cal H}_0$ to be defined by (\ref{quad}) and 
$\cal J$ by  (\ref{defJc}), with $R$ fixed to be either $r = +1$ or $r = -1$. The 
minimum eigenvalue of $\cal J$ is  still zero, but is now unique, the 
corresponding eigenvector being either $v_+$ or $v_-^{}$. 



To calculate the matrix elements $\tilde{Z}$, $\widetilde{W}$ using 
(\ref{trace}), we shall need to write each as the trace of an operator. We can 
do this by  introducing the operator   $  \exp(-\gamma {\cal J} )  $.
In the limit $\gamma \rightarrow +\infty$ we have
\be \label{grJ}
\e^{-\gamma {\cal J}}   \eq  v_r v_r^{\dagger}  \period \ee

It follows that 
\bd [{\cal H}_0, \Gamma_i ] = 2 \i \Gamma_{L+i+1} \sep 
[{\cal H}_0, \Gamma_{L+i}] = - 2 \i \Gamma_{i-1} \comma \ed

\be \label{commH}
 [{\cal H}_1, \Gamma_i ] = -2 \i \Gamma_{L+i} \sep 
[{\cal H}_1, \Gamma_{L+i}] = 2 \i \Gamma_{i} \comma \ee
for $i = 1, \ldots, L$, provided that on the RHS we take
$\Gamma_{2L+1} = - r \,  \Gamma_{L+1}$ and 
$\Gamma_0 = - r \, \Gamma_L$.

We shall need some sparse $L$ by $L$ matrices. Let $0$, $I$ be the 
zero and identity $L$ by $L$ matrices, respectively, and let
${\cal A}, {\cal B} $ be the one-off and two-off-diagonal $L$ by $L$
 matrices

\bd {\cal A}  \eq 
 \left( \begin{array}
{cccccccc} 0 & 1 & 0 &\cdot & \cdot &  \cdot  &\cdot\\ 
\cdot & 0 & 1 & 0 &  &   &  \\ 
\cdot  & & & \cdot & & &   \\ 
 \cdot   & & & &\cdot  &&    \\ 
 \cdot &  & &  &   0 & 1 & 0  \\ 
0 & \cdot  &  &  &   & 0 & 1  \\ 
- r & 0 &  \cdot &  \cdot &  \cdot &  \cdot & 0  \end{array}
 \right) \sep
 {\cal B}  \eq 
 \left( \begin{array}
{cccccccc} 0 & 0 & 1 & 0  & \cdot &  \cdot  &\cdot\\ 
\cdot & 0  & 0 & 1 &   0 &   &  \\ 
\cdot  & & & & \cdot  & &   \\ 
 \cdot   & & & & & \cdot  &    \\ 
0  &  & &  &  0  & 0 & 1  \\ 
- r   &  0  &  &  &   &  0  & 0  \\ 
0 & - r   &  0  &  \cdot &  \cdot &  \cdot & 0  \end{array}
 \right) \period \ed
Note that
\be {\cal A} {\cal A}^T = {\cal A}^T {\cal A} =  I  \sep {\cal B} = A^2 \period \ee

Then from (\ref{commH}),  the $H$-representatives are
 \be \widehat{\cal H}_0   \eq 2 \i 
 \left( \begin{array}
{cc} 0 & -{\cal A} \\  {\cal A}^T  & 0 \end{array} \right) \comma \ee

\be \widehat{\cal H}_1   \eq 2 \i 
 \left( \begin{array}
{cc} 0 & I \\ -I & 0 \end{array} \right) \period  \ee
 

\subsubsection*{Diagonalization of $\cal H$}


{From} (\ref{HHH}), the $H$-representative of $\cal H$ is
$\widehat{\cal H}_0 +k' \widehat{\cal H}_1$. 

We calculate this matrix.
First we define a $2L$ by$2L$ matrix $M$ 
(not to be confused with the $M$ of section 2):
\be M =  \left( \begin{array}
{cc} \i  & -\i {\cal A}  \\ {\cal A}^T & 1 \end{array} \right)  \ee (writing $\i$ 
for $\i  I$  and $1$ for $I$).
Then 
\be M^{-1} \widehat{{\cal H}}_0 M \eq  2
 \left( \begin{array}
{cc} -1  & 0 \\ 0 & 1 \end{array} \right) \comma \ee

\be M^{-1} \widehat{{\cal H}}_1 M \eq 
 \left( \begin{array}
{cc}  {\cal A} +  {\cal A}^T  & I-{\cal B}  \\ I-{\cal B}^T  & 
- {\cal A} - {\cal A}^T  \end{array} \right) \period  \ee
Thus $M^{-1} \widehat{\cal H}_0 M$ is diagonal.


For $j = 1, \ldots, L$, define
\ba \label{deftheta0}
\theta_j    =  &  \pi (2j-1)/L \; \; & {\rm if \; \;}  r = + \comma \nonumber \\
   =   & 2  \pi j /L \; \;  & {\rm if \; \;}  r = - \comma   \ea
   set
   \bd z_j \eq \e^{\i \theta_j } \comma \ed
   and let $\hat{P},  \hat{Q} $ be the $L$ by $L$ matrices with  entries
   \be \label{defP}  \hat{P}_{i,j} \eq     z_j^{i-1}/\sqrt{L}   \sep  \hat{Q}_{i,j} \eq  
      \i \, z_j^{i-2}/\sqrt{L}    \period \ee
   These matrices are unitary, but not orthogonal. Set 
   \be \label{defcalP}
   {\cal P} \eq  
   \left( \begin{array} {cc} 
   \hat{P}  & 0  \\ 0  & \hat{Q}  \end{array} \right) \period  \ee
Then \be \label{Htilde}
\tilde{\cal H}_0 \eq {\cal P}^{-1} M^{-1} \widehat{\cal H}_0 M {\cal P}
\eq 
 2 \left( \begin{array}
{cc}  -1  & 0  \\ 0 & 1  \end{array} \right) \comma  \ee
\bd
\tilde{\cal H}_1 \eq {\cal P}^{-1} M^{-1} \widehat{\cal H}_1 M {\cal P}
\eq 
2 \left( \begin{array}
{cc}  \tilde{C}  & \tilde{S}  \\ \tilde{S} & -\tilde{C}  \end{array} \right) 
\comma  \ed
where $\tilde{C}$, $\tilde{S}$ are diagonal $L$ by $L$ matrices with 
diagonal entries
\be \tilde{C}_{j,j} = \cos \theta_j \sep \tilde{S}_{j,j}  = \sin \theta_j \period 
\ee


   Thus $\tilde{\cal H} = \tilde{\cal H}_0 +k' \tilde{\cal H}_1$ can 
be re-arranged as a matrix consisting of $L$ two-by-two diagonal 
blocks. It is then straightforward to calculate its exponential, giving 
\be \label{Hct}
\exp(-\alpha {\tilde{\cal H}}) \eq  
 \left( \begin{array}
{cc}  \tilde{U}  & \tilde{V}  \\ \tilde{V} & \tilde{T}  \end{array} \right)
 \comma  \ee
where $\tilde{U} =  \tilde{U}_r(\alpha), \, \tilde{V} = 
 \tilde{V}_r(\alpha), \,  \tilde{T} =  \tilde{T}_r(\alpha)$ are diagonal matrices
with diagonal entries (for  $j = 1,\ldots, L$)
\bd 
\tilde{U} _{jj} =  u_j(r, \alpha) = \cosh(2 \alpha \lambda_j) 
+ \frac{1-k' \cos \theta_j }{\lambda_j} \sinh(2 \alpha \lambda_j ) \comma \ed
\be  \label{uvt}
\tilde{V} _{jj} =  v_j(r, \alpha) = -\frac{k' \sin \theta_j 
\sinh( 2\alpha \lambda_j)}{\lambda_j }   \comma \ee
 \bd \tilde{T} _{jj} =  t_j(r, \alpha) = \cosh(2 \alpha \lambda_j) 
- \frac{1-k' \cos \theta_j }{\lambda_j} \sinh(2 \alpha \lambda_j )  \ed
and
\be \label{deflam}
\lambda_j \eq  \left( 1 - 2 k' \cos \theta_j +k'^2 \right)^{1/2}
\period \ee



\subsubsection*{Calculation  of $\tilde{Z}$}


We want to calculate $\tilde{Z}_+(\alpha)$ and $\tilde{Z}_-^{}(\beta)$ from 
(\ref{defZ}). First consider $\tilde{Z}_+(\alpha)$, so take $r = +$ in the 
above equations. If we are working in the 
sub-space ${\cal V}_+$,
we can use (\ref{grJ}) to write
(\ref{defZ}) as 
\be  \tilde{Z}_{+} (\alpha) \eq \trace  \e^{-\gamma {\cal J}}
 \e^{-\alpha {\cal H}}    \ee
in the limit $\gamma \rightarrow + \infty$. Since ${\cal H}_0$ and 
${\cal H}_1$
are traceless,
\be \det \e^{-\rho {\cal H}_0}= \det \e^{-\alpha {\cal H}} = 1 \ee
and from (\ref{defJc}),  $\det \e^{-\gamma {\cal J}} = \e^{-L d \gamma }$,
$\widehat{\cal J} = \widehat{\cal H}_0$. {From}   (\ref{trace}) 
we therefore  have
\be  \label{Zt}
\e^{2 L \gamma}  
\tilde{Z}_{+} (\alpha)^2  \eq  \det ( I + \e^{-\gamma \widehat{\cal H}_0}
 \e^{-\alpha \widehat{\cal H}}  )  =
 \det (  \e^{\gamma \widehat{\cal H}_0} +
 \e^{-\alpha \widehat{\cal H}}  ) \period \ee
 Using the similarity transformation of (\ref{Htilde}), we can replace
 $\widehat{\cal H}_0$,  $\widehat{\cal H}$ in (\ref{Zt}) by
  $\tilde{\cal H}_0$,  $\tilde{\cal H}$, giving
  \be \label{Ztsq}
\e^{2 L \gamma}  
\tilde{Z}_{+} (\alpha)^2  \eq 
\det  \left( \begin{array}
{cc}  \e^{-2\gamma} I + \tilde{U}  & \tilde{V}  \\ \tilde{V} & \e^{2 \gamma}I  +
\tilde{T}  \end{array} \right)   \period  \ee
 In the limit of $\gamma$ large we expect the $ \e^{-2\gamma} I $, 
 $\tilde{V}$, $\tilde{T}$   blocks in this determinant to become relatively 
 negligible. The factors involving $\gamma$ then cancel,  leaving
 \be \label{Ztsq2}
 \tilde{Z}_{+} (\alpha)^2   =  \det  \tilde{U}   =
 \det  \tilde{U}_+(\alpha)    = \prod_{j=1}^L u_j(+,\alpha)  \period \ee

Similarly,
\be \label{Ztsq3}
 \tilde{Z}_{-} (\alpha)^2   =  
 \det  \tilde{U}_-^{}(\alpha)    = \prod_{j=1}^L u_j(-,\alpha)  \period \ee
 

\subsubsection*{Calculation  of $\widetilde{W}$}


The function $\widetilde{W}(\alpha, \beta, x)$ is defined by (\ref{defW}). 
We note that the operator $S_1$ takes a vector in 
${\cal V}_+$ to one in ${\cal V}_-^{}$, and vice-versa. In particular,
\bd  S_1  v_r^{} \eq v_{-r}^{}  \ed
so 
\be \lim_{\gamma \rightarrow + \infty}    \e^{-\gamma {\cal J}} S_1
=   v_{-r}^{} v_r^{\dagger}  \period \ee

We can use this to write (\ref{defW}) as a trace:
\be  \label{defW2}  
 \widetilde{W} (\alpha,\beta, x) \eq  \trace 
\e^{-\gamma {\cal J}}    S_1 
   \e^{-\alpha {\cal H}} 
 \e^{-\rho {\cal J}} S_1
 \e^{-\beta {\cal H}}   \comma  \ee
in the limit $\gamma \rightarrow + \infty$.
    We take the factors $\e^{-\gamma {\cal J}} $, $\e^{-\beta {\cal H}} $
 to be acting in ${\cal V}_-^{}$, the two other exponential
 factors to be acting in ${\cal V}_+^{}$.
 
 The reason we are able to make further progress is that
 $S_1$ also belongs to the group $\cal G$. In fact 
 $S_1 = \Gamma_{L+1}$. This is linear in the $\Gamma_j$, unlike
 ${\cal H}_0$ and  ${\cal H}_1$, which from (\ref{quad}) are
 quadratic. Even so,
 \be 
 S_1 \Gamma_j S_1^{-1} =  \epsilon_j \Gamma_j \comma \ee
 where $\epsilon_{L+1} = +1 $, else   $\epsilon_j  = -1 $. 
 The representative  of $S_1$
  is therefore 
  \be \widehat{S}_1  = \left( \begin{array}
{cc} -I  & 0 \\ 0 & -  E \end{array} \right) \comma \ee
where $E$ is the diagonal matrix
\be E =  \left( \begin{array}
{cccccccc} -1 & 0  &\cdot & \cdot &  \cdot  &\cdot\\ 
 0 & 1 & 0 &  &   &  \\ 
 \cdot   & &  &\cdot  &&    \\ 
 \cdot &  &  &   0 & 1 & 0  \\ 
\cdot  & \cdot  &  \cdot &  \cdot &  \cdot & 1  \end{array}
 \right)  \period \ee
 
  
  Remembering that $\widehat{\cal J} = \widehat{\cal H}_0$,
  the representative of the matrix whose trace is to be evaluated
  in (\ref{defW2}) is therefore the $2L$ by $2L$ matrix
  \be
 \widehat{\cal  W}(\alpha, \beta, x) =   \e^{-\gamma { \widehat{\cal H}_0}}
 \widehat{S}_1   \e^{-\alpha {   \widehat{\cal H}}} 
 \e^{-\rho    \widehat{{\cal H}_0}}    \widehat{S}_1
 \e^{-\beta {   \widehat{\cal H}}}   \ee
 and from (\ref{trace})
 
 
 \be \label{ Wdet}
 \e^{2 \, L (\gamma+\rho)}  \widetilde{W} (\alpha,\beta, x)^2 \eq 
 \det  \left[ { I} +  \widehat{ W}(\alpha, \beta, x) \right] 
 \period \ee

  We now use the similarity transformation of (\ref{Htilde}). We 
 have to be careful because $A,  M, \hat{P}, \hat{Q},  {\cal P}$ 
 depend on $r$. We write them more explicitly as  
 $A_r,  M_r,  \hat{P}_r, \hat{Q}_r, {\cal P}_r$. 
Similarly we 
write the  $\theta_j$, $z_j$ of   (\ref{deftheta0}) - (\ref{defcalP}) as
$\theta_{r,j}$,  $z_{r,j}$.
For the middle two exponential factors we take $r=+$, for the others
 we take $r = -$. The result is
 \be \label{Wt} 
 \tilde{\cal W} = {\cal P}_-^{-1} M_-^{-1} \widehat{\cal W} M_-^{} {\cal P}_-^{} 
 =  \e^{-\gamma\tilde{\cal H}_0}
 \tilde{S}_-^{}
   \e^{-\alpha {   \tilde{\cal H}}} 
 \e^{-\rho    \tilde{{\cal H}_0}}    \tilde{S}_+
 \e^{-\beta    \tilde{\cal H}}  \comma \ee
where 
\be \label{defSt}
 \tilde{S}_r \eq   {\cal P}_r^{-1} M_r^{-1} 
 \widehat{S}_1 M_{-r}^{} {\cal P}_{-r}^{} \period  \ee
 
 This threatens to become messy, but we find some remarkable
 simplifications. Firstly,
 \be M_r \widehat{S}_1 = \widehat{S}_1 M_{-r} \comma
 \ee
 so $ M_r^{-1}  \widehat{S}_1 M_{-r}^{} =  \widehat{S}_1$
 and (\ref{defSt}) becomes
 \be 
 \tilde{S}_r \eq   {\cal P}_r^{-1} 
 \widehat{S}_1 {\cal P}_{-r}^{} \period  \ee
 
 The second surprise is that
 \be  \label{FPQ}
 \hat{Q}_r^{-1} E \, \hat{Q}_{-r} = 
 \hat{P}_r^{-1}  \hat{P}_{-r} =  - F_r  \comma \ee
 where $F_r$ is the $L$ by $L$ matrix with entries
 \be \label{defF}
 F_{i,j}^{(r)} = \frac{2 \, z_{r,i} }{ L ( z_{-r,j} - z_{r,i}) } \comma  \ee
 satisfying 
 \be \label{FFprod}
 F_-^{} F_+ = I \period \ee
 Hence $ \tilde{S}_r $ has the comparitively simple block-diagonal 
  form
 \be
 \tilde{S}_r \eq \left( \begin{array}
{cc} F_r  & 0 \\ 0 &  F_r \end{array} \right) \period \ee


Let \be {  \tilde{Y}} \eq  \tilde{S}_-^{}
   \e^{-\alpha {   \tilde{\cal H}}} 
 \e^{-\rho    \tilde{{\cal H}_0}}    \tilde{S}_+
 \e^{-\beta    \tilde{\cal H}}   \comma \ee
 then from (\ref{ Wdet}) and (\ref{Wt})
 \be 
  \e^{2 \, L (\gamma+\rho)}  \widetilde{W} (\alpha,\beta, x)^2 \eq 
    \det \left(  I + \e^{-\gamma {\tilde{\cal H}}_0}  \tilde{Y}  \right)
   \eq   \det \left( \e^{\gamma {\tilde{\cal H}}_0} I +   \tilde{Y}  \right) \period
   \ee
 The RHS has the same stucture as that of (\ref{Ztsq}), so taking
 the limit $\gamma \rightarrow + \infty$, we obtain, similarly to 
   (\ref{Ztsq2}),
   \be    \e^{2 \, L \rho}  \widetilde{W} (\alpha,\beta, x)^2 \eq 
   \det  \tilde{Y}_{11} \comma \ee
 where $ \tilde{Y}_{11}$ is the top-left $L$ by $L$ block of $ \tilde{Y}$.
 Using (\ref{Htilde}) and (\ref{Hct}), it follows that
 \be  \label{resW}
  \widetilde{W} (\alpha,\beta, x)^2 \eq 
     \det \left[ F_-^{} \tilde{U}_+(\alpha) \, F_+ \tilde{U}_-^{}(\beta) +
    x  F_-^{} \tilde{V}_+(\alpha) \, F_+ \tilde{V}_-^{}(\beta)  \right] \period \ee

 Set, for $r = + $ or $-$,
  \be \label{defX}
  X_{r}(\alpha ) =  \left[  \tilde{U}_{r} (\alpha) \right]^{-1}  
 \tilde{V}_{r}(\alpha) 
 \period  \ee
 Then from  (\ref{Ztsq2}) and (\ref{Ztsq3}), remembering that the $ \tilde{U},  
 \tilde{V}$ matrices are diagonal, so commute,
 we can write (\ref{resW}) as
\be   \label{resW2}
  \widetilde{W} (\alpha,\beta, x)^2 \eq 
  \tilde{Z}_{+}(\alpha)^2 \tilde{Z}_{-}^{}(\beta)^2    \det \left[  I+
    x  F_+^{-1} X_+(\alpha) \, F_+ X_-^{}(\beta)  \right] \period \ee

To summarize thus far, the equations (\ref{Ztsq2}), 
(\ref{Ztsq3}), (\ref{resW2}) give $\tilde{Z}_{+}(\alpha)$,
 $\tilde{Z}_{-}(\beta)$, $  \widetilde{W} (\alpha,\beta, x)$ in terms of
  $L$ by $L$ determinants, and these 
 formulae are exact for finite  $\alpha, \beta, L$.  The 
 spontaneous magnetization $\cal M$ is then given by 
 (\ref{resM}), taking the limit $\alpha, \beta, L \rightarrow +\infty$.



 \section{Calculation of ${\cal M}$}
 
 \setcounter{equation}{0}

 So far we have parallelled the method of Yang.\cite{Yang1952} We now 
explicitly calculate  $\cal M$ from the above equations, but
we  use a method more like that of Montroll, Potts and 
Ward.\cite{MPW1963}
 
 First set $x=1$ and take the limit $\alpha, \beta \rightarrow + \infty$ in 
 (\ref{Ztsq2}), 
(\ref{Ztsq3}), (\ref{resW2}), using (\ref{uvt}). We obtain
\be \label{Mr1}
{\cal M}^2 \eq \xi_{+}  \xi_-^{}  \det \left[  I+
     F_+^{-1} X_+(\infty) \, F_+ X_-^{}(\infty)  \right] \comma \ee
where
\be \xi_r \eq \prod_{j=1}^L\left(  \frac{\lambda_{r,j}+1-
k' \cos \theta_{r,j}}{2 \lambda_{r,j} } \right)^{1/2} \comma \ee
and $X_r(\infty) $ is the diagonal matrix with diagonal entries
\be 
\left[X_r(\infty)  \right] _{j,j} \eq \frac{ - k' \sin \theta_{r,j} }{
\lambda_{r,j} +1 - k' \cos \theta_{r,j}} \period \ee


 {From} (\ref{FPQ}), $F_+ = - {\hat{P}}_+^{-1} \hat{P}_-^{}$, 
 so if we set
 \be 
 \tilde{X_r} \eq \hat{P}_r X_r (\infty) \hat{P}_r^{-1} \ee
 then (\ref{Mr1}) can be written
 \be \label{Mr2}
{\cal M}^2 \eq \xi_{+}  \xi_-^{}  \det \left[  I+
 \tilde{X}_+  \tilde{X}_-^{}   \right] \period \ee
 
 We could write this determinant as
 \bd \det  \tilde{X}_+ \; \det  \left[  
 \tilde{X}_+^{-1}+   \tilde{X}_-^{}   \right] \period \ed
 {From} the definition (\ref{defP}) of $\hat{P}$, the matrices 
 $ \tilde{X}_+,   \tilde{X}_-^{} $ are anti-cyclic and cyclic, 
 respectively, i.e.
 \bd [\tilde{X}_r]_{i,j} =  \eta_{i-j} = - r \, \eta_{L+i-j}  \comma \ed
 where $\eta$ is defined by this equation. The matrix 
 $ \tilde{X}_+^{-1}+   \tilde{X}_-^{} $ is therefore Toeplitz, its elements
 $i,j$ depending only on $i-j$, and one might hope to use 
 Szeg{\H o}'s theorem\cite{MPW1963,Szego1958} to evaluate the 
 determinant for large  $L$.
 
 Unfortunately this naive approach does not work because 
 for large $L$ the determinant is not dominated by its near-diagonal 
 elements, in particular elements such as $(L,1)$, $(2,L \minus 3)$
 do not tend to zero. However, we can rescue this idea by making
 a particular bilinear transformation of  $ \tilde{X}_+,   \tilde{X}_-^{} $.
 
 We set
 \be \label{defC}
 C_r \eq (I-\i  \tilde{X}_r )^{-1} (I+\i  \tilde{X}_r )
 \eq \hat{P}_r c_r  \hat{P}_r^{-1} \comma \ee
 where $c_r$ is a diagonal matrix with diagonal elements
 \be (c_r)_{j,j}  = \frac { \lambda_{r,j} +1 -k' \exp(\i \theta_{r,j})}
 { \lambda_{r,j} +1 -k' \exp(-\i \theta_{r,j})} \period \ee
 Solving for $X_r$ and substituting into (\ref{Mr2}), we obtain
 \be \label{Mr3}
 {\cal M}^2 \eq \xi'_{+}  \xi'_-  \det  [ (C_+  +
 C_-^{} )/2 ] \comma \ee
 where
 \ba   \xi'_r & = &  \xi_r \det (I - \i \tilde{X}_r ) \nonumber \\
 & = & \prod_{j=1}^L \left\{ \frac{[\lambda_{r,j} + 1 - k' \exp(-\i \theta_{r,j})]^2}
 {2 \, \lambda_{r,j} (\lambda_{r,j}+1 -k' \cos \theta_{r,j}) } \right\}^{1/2}
 \period \ea

 {From} (\ref{deflam}),
 
 \bd \lambda_{r,j}^2 =  [1-k' \exp ( \i \theta_{r,j})]  \,  [1-k' \exp(- \i \theta_{r,j})]
 \comma \ed
 from which it follows that
 \be  \xi'_r  =  \prod_{j=1}^L\left[  \frac { 1-k' \exp(- \i \theta_{r,j})}
 { 1 -k' \exp(\i \theta_{r,j})} \right]^{1/4} \period 
 \ee
 The $\theta_{r,j}$ are either $\pi$ or $2 \pi$, or occur in pairs
 $\theta_{r,j}, 2 \pi - \theta_{r,j}$. It follows that
 \be \xi'_+=  \xi'_-= 1 \period \ee
 
 Also, $(c_r)_{j,j}$ simplifies to
 \be (c_r)_{j,j}  \eq \left[  \frac { 1-k' \exp(- \i \theta_{r,j})}
 { 1 -k' \exp(\i \theta_{r,j})} \right]^{1/2} \period \ee

  {From} (\ref{defP}) and  (\ref{defC}), the elements of $C_r$ are
  \be   \label{elC}
  (C_r)_{i,j} \eq \frac{1}{L} \sum_{m=1}^L (c_r)_{m,m} z_{r,m}^{i-j}
  \comma \ee
  where $z_{r,m} = \e^{\i \theta_{r,m}}$. The summand depends on $m$
 via $\theta_{r,m}$ and is an analytic function of $\theta_{r,m}$ on the real 
 axis, periodic of period $2 \pi$. The  $\theta_{r,m}$ are distributed uniformly
 throughout this period. If $i,j$ are held finite and $L \rightarrow \infty$, it 
 follows that 
 \be  \label{limC}
  (C_r)_{i,j} \eq \frac{1}{2 \pi } \int_{0}^{2 \pi}  f(\theta) \e^{\i (i-j) \theta} 
 {\rm d}  \theta \comma \ee
 where
 \be  \label{dff}
   f(\theta) \eq  \left[  \frac { 1-k' \exp(- \i \theta)}
 { 1 -k' \exp(\i \theta)} \right]^{1/2}  \period \ee
 
 This is actually the function $\e^{\i \delta^*}$ of 
 Onsager\cite[eq. 89]{Onsager1944} and 
 Montroll {\it et al}\cite[eq. 42]{MPW1963}  In their notation $z_1   = \tanh H'$,
 $z_2^* = \tanh H^*$ and our $k'$ is  $k' = z_2^*/z_1$. We have replaced  
 $\omega$ by 
 $\theta$ and taken the hamiltonian
  limit $z_1, z_2^* \rightarrow 0$.
 
 Note that the limit (\ref{limC}) is the same for $r = +1$ and $r = -1$. Also,
for large but finite $L$, the corrections to  (\ref{limC})  vanish
exponentially with $L$. The near-diagonal elements of $C_+$ and 
$C_-^{}$ therefore become equal when $L$ is large.
   
   Since $z_{r,m}^L = -r$,  incrementing $i$ or $j$ in (\ref{elC}) by $L$
negates $  (C_r)_{i,j}$ for $r=+$, and leaves it unchanged for $r=-$.
This means that the near-top-right and near-bottom-left elements of 
$C_+$ and  $C_-^{}$ also tend to finite limits as $L$ becomes large, but are 
equal and opposite. Their sum  therefore approaches exponentially
to zero .

Hence when $L$ is  large we expect (\ref{Mr3}) to become
\be {\cal M}^2 \eq \det C \comma \ee
where $C$ is the $L$ by $L$ matrix with elements $(i,j)$ given by
(\ref{limC}). These elements tend exponentially to zero as $|i-j|$ becomes
 large.   This is a special case of the Toeplitz matrix
discussed in \cite{MPW1963} and we can use the
general result (68) therein:
\be \lim_{L \rightarrow \infty}  G^{-L}   \,  \det C   \eq \exp\left( 
\sum_{n=1}^{\infty} n \kappa_n \kappa_{-n} \right) \comma \ee
where 
\bd G = \exp \left( \frac{1}{2\pi} \int_0^{2 \pi} 
\log  f(\theta ) {\rm d} \theta \right) \comma \ed
and \bd \log f (\theta ) \eq \sum_{n=-\infty}^{\infty} \kappa_n \e^{\i n \theta } 
\period \ed

Since $\log f( \theta)$ is an odd periodic function of $\theta$, we have 
$G=1$. Also, taking the logarithm of (\ref{dff}) and Laurent expanding,
we readily find
\be
2 \kappa_n =   {k'}^{\, n}/n \; \; {\rm if \;} n >0 \sep 2 \kappa_n = 
 - {k'}^{\, -n} /n\; \; {\rm if \;} n < 0   \period \ee
 It follows that
 $ \det C \eq (1-k'^2)^{1/4} $,
 and hence
 \be {\cal M } \eq (1-k'^2)^{1/8} \period \ee
 This is of course Onsager's famous 
 result.\cite{MPW1963, Onsager1949, Yang1952}.
 

 


 \section{Connection with the superintegrable chiral Potts model}
 

\setcounter{equation}{0}

The superintegrable case of the $N$-state chiral Potts model has some 
properties that
closely resemble the Ising model. In particular, there is a spin-shift operator
$R$ that is the natural generalization of (\ref{defR}) and divides the vector 
space into $N$ sub-spaces ${\cal V}_{\Q}$, labelled by 
${\Q} = 0, 1, \ldots N_1$.  
Within  each ${\cal V}_{\Q}$, if  one imposes the fixed-spin 
boundary conditions of Fig. 1, the transfer matrix $T$ generates a yet 
smaller sub-space of dimension $2^m$,
where
\be \label{defm} m = m_{\Q} =  
 \left[\frac{(N-1) L - {\Q}}{N} \right] \period \ee
Here $[x]$ means the integer part of the real number $x$.

 Within  ${\cal V}_{\Q}$, $T$ is a direct product of $m$ two-by-two matrices, 
 and  there are similarity transformations that reduce the associated 
 hamiltonian  to the direct  sum\cite[eq.2.20]{Baxter1989}
 \bd
{\cal H} \eq \mu_{\Q}- N \sum_{j=1}^m [(1-k' \cos \theta_j ) S_j - 
k' \sin \theta_j \, C_j ] \comma \ed
 where
\be \mu = \mu_{\Q} = 2 k' {\Q} +(1+k')(mN-NL+L) \ee
and  $\theta_1, \ldots \theta_m$ are defined by
\be \label{deftheta}
\cos \theta_j \eq  (1+w_j)/(1-w_j) \sep 0 
< \theta_i < \pi  \comma \ee
the $w_1, \ldots ,w_m $  being the zeros of 
\be \label{defpolP}
 P(z^N) \eq z^{-{\Q}} \, \sum_{n=0}^{N-1} 
\omega^{(L+{\Q})n} {[(z^N-1)/(z-\omega^n)]}^L \comma \ee
which is  a polynomial in $w = z^N$ of degree $m$. 

 
\subsection*{The partition functions $\tilde{Z}$}

One can explicitly calculate\cite{Baxter2008b} the  partition 
functions 
that generalize (\ref{Ztsq2}), (\ref{Ztsq3}):
 \be \label{resvq}
 {\tilde{Z}}_{\Q} (\alpha)  \eq e^{-\mu \alpha} 
\, u_1(\alpha) \cdots u_m(\alpha ) \comma 
\ee
the function $u_j(\alpha) $ again being defined by 
(\ref{uvt}), (\ref{deflam}).

 For $N= 2$ the model reduces to the Ising model with
 \be r= 1-2 {\Q}   \sep   m(r)  = m_{\Q} \comma \ee
 so ${\Q}=0$  corresponds to $R= +1$,
 while ${\Q}=1$ corresponds to $R= -1$.
 The result (\ref{resvq}) of course
 agrees with (\ref{Ztsq2}), (\ref{Ztsq3}), but has a slightly different form.
  The partition function is no longer squared on the LHS and 
  instead of there being $L$ variables $\theta_{j}$, there are only $m$, 
 which lie between $(L-2)/2$ and $L/2$. They are the same as the 
  $\theta_{j}$ of the previous sections, but $j$ takes only the values 
  $1$ to $m$ and  $0 < \theta_{j} < \pi$.
  

  These differences are easily explained. For every  $\theta_{j}$
  with $j = 1 , \ldots , m$ there is a $\theta_{j'} = 2\pi - \theta_j$, where
$j' = L+1-{\Q}-j $ and $m < j' \leq L$. They have the same value of 
 $u_j(\alpha)$ and  each occurs in  (\ref{Ztsq2}), (\ref{Ztsq3}),
while only one occurs in (\ref{resvq}). This accounts for the absence of the
square in (\ref{resvq}) and the presence of the factor
$e^{-\mu \alpha} $ accounts for the exceptional cases
$\theta_j  = \pi $ or $2 \pi$, which are included in the product in
 (\ref{Ztsq2}), (\ref{Ztsq3}).
 
 There are a total of four cases to consider.
 
 1) $r$ = +1, ${\Q} = 0$, $L = $ even: there are no exceptional cases
 and $m=L/2$, $\mu = 0 $.
 
 2)  $r$ = +1, ${\Q} = 0$, $L = $ odd: then $m = (L-1)/2, \mu = -1-k'$.
 There is an exceptional case at $j = (L+1)/2 = m+1$, where
 $\theta_j = \pi$, $u_j(\alpha) = \exp[2 (1+k') \alpha] = \e^{-2 \mu \alpha}$.

 3)  $r$ = -1, ${\Q} = 1$, $L = $ even: then $m = (L-2)/2, \mu = -2$.
 There are two exceptional cases at $j = L/2$ and $ L$, with 
 $\theta_j = \pi$ and $2\pi$,
 $u_j(\alpha) =  \exp[2 (1+k') \alpha] $ and $  \exp[2 (1-k') \alpha] $, so
 $u_{L/2}(\alpha) \, u_L(\alpha) = \e^{-2 \mu \alpha}$.

4)  $r$ = -1, ${\Q} = 1$, $L = $ odd: then $m = (L-1)/2, \mu = k'-1$.
 There is an exceptional case at $j=L$, with $\theta_j = 2\pi$,
  $u_j(\alpha) =  \exp[2 (1-k') \alpha]  = \e^{-2 \mu \alpha}$.
  
  In each case the total contribution of the exceptional cases to 
 the RHS of  (\ref{Ztsq2}) or (\ref{Ztsq3}) is $\e^{-2 \mu \alpha}$, which 
 accounts for the factor $e^{-\mu \alpha} $ in (\ref{resvq}).

 \subsection*{The partition function $\widetilde{W}$}
 

 The equation  (\ref{resvq}) simplifies  (\ref{Ztsq2}) and (\ref{Ztsq3}), 
 expressing
 the $\tilde{Z}_{\Q}(\alpha)$ as a product rather than the square root of a 
 product,  and reducing the number of factors from $L$ to $m$ or $m'$,
where
\be \label{mmp}
 m = m(+) = [L/2] \sep m' = m(-) = [(L-1)/2] \period \ee
 Can we similarly
 reduce the expression (\ref{resW}) for $\widetilde{W}(\alpha, \beta, x)$?
 
 The answer is yes, apart from simple factors that are independent of $x$
 and easily calculated. We show that  the matrix
 \be \label{defGr}
 G = \tilde{U}_+ (\alpha) \, F_+  \tilde{U}_{-}^{}(\beta) +
    x   \tilde{V}_+ (\alpha) \, F_+  \tilde{V}_{-}^{}(\beta)   \ee
  can be reduced to block lower-triangular form and its determinant
 expressed as simple factors times the square of an $m'$ by 
 $m'$ determinant.


 For brevity, in the following two sub-sections, unless 
 indicated otherwise  we write
 \bd \theta_j = \theta_{+,j} \sep \theta'_j = \theta_{-,j} \comma \ed
 \bd u_j = \tilde{U}_+(\alpha)_{j,j} = u_{+,j}(\alpha) \sep 
 v_j = \tilde{V}_+(\alpha)_{j,j} = v_{+,j}(\alpha) \comma \ed
 \bd u'_j = \tilde{U}_-^{}(\beta)_{j,j} = u_{-,j}(\beta) \sep 
 v'_j = \tilde{V}_-^{}(\beta)_{j,j} = v_{-,j}(\beta) \period \ed
 
 We use the definition  (\ref{defF}) of $F_{\pm}$ 
 and the properties (\ref{FFprod}), (\ref{refF}).  We also
 implicitly use
 \be u_{L+1-j} = u_j \sep v_{L+1-j} = - v_j  \sep 
 u'_{L-j} = u'_j  \sep v'_{L-j} = - v'_j  \period \ee
 We do {\em not } in this section need the definitions  (\ref{uvt}).
 
 
 \subsubsection*{The case $L$ even}
 
 In this case note that the $\tilde{U}_+(\alpha) $,  $\tilde{V}_+(\alpha) $
 on the left of (\ref{defGr}) depend on the $\theta_{j}$, which satisfy
  $\theta_{i} +  \theta_{L+1-i} = 2 \pi$, so it is natural to combine
  rows $i$ and $L+1-i$. The  $\tilde{U}_{-}^{}(\beta) $,  $\tilde{V}_{-}^{}(\beta) $
  on the right involve $\theta'_{j}$, satisfying   $\theta'_{j} +  
  \theta'_{L-j} = 2 \pi$, so it is natural to combine
  columns  $j$ and $L-j$. Also, columns $L/2$ and $L$ both 
  correspond to exceptional values of $\theta'_j$, where $\sin \theta'_j = 0$.
  The elements $v'_{L/2}, v'_L$ of $\tilde{V}_{-}^{}(\beta)$ vanish 
  for these two columns.  {From} (\ref{mmp} ), 
 \be m =  L/2 \sep m'  = (L-2)/2 = m-1 \period \ee.


  We perform the  following equivalence transformations sequentially on
  $G$:
  \ba  
 \label{6steps}
 1)  \; \; G_{ij} \rightarrow  &  G_{ij} + G_{L+1-i,j}  ,   & 
 1\leq i \leq m, \; 1 \leq j \leq L \nonumber \\
    2)  \; \; G_{ij}   \rightarrow  & G_{ij} + G_{i,L-j}  ,  &  m < j < L  
     \nonumber \\
3)  \; \;  G_{i,L}   \rightarrow  \! \!   & G_{i, L} -  u'_L G_{i,m}/u'_{m}  , 
  &   \nonumber \\
 4)  \; \; G_{i,j}   \rightarrow  & G_{i, j} -  u'_j G_{i,m}/u'_{m}  ,  
 &  1 \leq j < m   \\
5)  \; \;   G_{i,j}   \rightarrow  & G_{i, j} - 2 u'_j  G_{i,m}/u'_m, 
&  m < j < L  \nonumber \\
6)  \; \;  G_{i,j}    \rightarrow  & G_{i, j} - (1+ c'_j)
\, u'_j  G_{i,L}/u'_L,  & m < j < L   
  \nonumber \ea
  where $c'_j = \cos ( \theta'_{j})$.

The first transformation corresponds to pre-multiplying $G$  by 
some matrix, all the others 
to post-multiplying it. Steps (2) to (6) are for  all values of $i$, 
i.e. $1\leq i \leq  L$.  At every step we are merely incrementing rows or
columns by linear combinations of other rows or columns, so the 
determinant of $G$ is unchanged.


The elements $F_{i,j}$ of $F_+$ satisfy
  \be \label{refF}
  F_{ij} +   F_{L+1-i,L-j}  = F_{i,L} + F_{L+1-i,L} = -2/L \ee
  for all $i,j$. It follows that the final matrix $G$ is such that
  \be G_{i,j} = 0 \; \;  {\rm for \; \; } 1\leq i \leq m, \;  \; m < j \leq L \period \ee

Thus $G$ is now block lower-triangular:
\be \label{valG}
G = \left( \begin{array}
{cc} Y  & 0 \\ {\cal T} & {\cal Z}\end{array} \right) \comma \ee
where the   $Y , {\cal T}, {\cal Z}$ are $m$ by $m$ matrices.
We find
\be \label{defY1}
Y_{i,j} \eq  \frac{- 2 \i \, (\sin\theta'_j  u_i u'_j +
x \sin \theta_i v_i v'_j )}
	{L( \cos \theta_i - \cos \theta'_j )}   \ee
\bd Y_{i,m} \eq -2 u_i u'_m/L  \ed
for $ 1\leq i \leq m , 1 \leq j < m$. 

Note that the equivalence transformations are independent of $x$, so
$Y, {\cal T}, {\cal Z}$ are all linear in $x$, their coefficients being the 
transforms of the corresponding coefficients of $G$.

We define the $m$ by $m$ matrix $\cal Z$ by (\ref{valG}) , but with the 
rows and columns
re-arranged so that, for $1 \leq i,j \leq m$,
\be \left(  {\cal Z} \right) _{i,j} \eq \left(  G \right) _{L+1-i,j'} \comma \ee
where $j' = L$ if $j=m$, else $j'= L-j$. Let $D,D'$ be the diagonal matrices
with elements
\be \label{gdD}
D_{j,j} = \sin\theta_i \sep D'_{j,j} = \sin\theta'_i \ee
for $j=1 , \ldots, m$, except that $D'_{m,m} = -\i u'_L/u'_m$. Then we find that
\be {\cal Z}  \eq - D^{-1} Y D' \period \ee
Hence 
\be \label{detG}
\det G = \det ( - D') (\det Y)^2/  \det D   \ee
and we see that $\det G$ is basically the square of the $m$ by $m$
determinant $\det Y$.


We can do better yet and write it in terms of the square of an $m' =m-1$ 
determinant.  Expand the matrices $G, Y$ in powers of $x$:
\be \label{expand}
 G = {\bf g} _0 + x {\bf g} _1 \sep Y  = {\bf y} _0+ x {\bf y}_1 \comma  \ee
so    ${\bf g}_0, {\bf g}_1,  {\bf y}_0, {\bf y}_0$ are  $x$-independent matrices
and ${\bf g}_0  =   \tilde{U}_+ (\alpha) \, F_+  \tilde{U}_{-}^{}(\beta)  $.
 {From} (\ref{FFprod}), its inverse is
\be   {\bf g}_0 ^{-1}  \eq   \tilde{U}_{-}^{} (\beta)^{-1}  \, F_{-}^{} 
\,  \tilde{U}_{+}^{}(\alpha) ^{-1} \period \ee
We can follow the above six steps (\ref{6steps}) and evaluate 
${\bf g}_0^{-1} $ after the six equivalence transformations. It also 
becomes block lower-triangular, as in (\ref{valG}). In particular its
top-left block is ${\bf y}_0^{-1}$, with elements
\be \label{Y10}
 \left[ {\bf y}_0^{-1} \right]_{i,j} \eq \frac{-2 \i \sin \theta'_i}
{ L (\cos \theta'_i - \cos \theta_j ) u'_i u_j } \comma \ee
\bd   \left[   {\bf y}_0^{-1} \right]_{m,j}  \eq -1/(u'_m u_j ) \comma \ed
for $1 \leq i < m, 1 \leq j \leq m$. This matrix is necessarily the inverse of 
${\bf y}_0$ and from (\ref{detG})
\be  \label{detG2}
\det {\bf g}_0^{-1} G \eq [ \det   {\bf y}_0^{-1} Y ]^2 \period \ee

We calculate ${\bf y}_0^{-1} Y$. {From} (\ref{defY1}),  {\em the last 
column of ${\bf y}_1$ vanishes.} Hence, using (\ref{expand}), 
\be  {\bf y}_0^{-1} Y \eq I +x \tilde{\cal D} \comma \ee
where $\tilde{\cal D}  =  {\bf y}_0^{-1} {\bf y}_1$ and again the last 
column of $\cal D$ vanishes. The determinant of
the RHS is therefore the same as that of the RHS {\em truncated to its 
first $m' = m-1$ rows and columns.}

We can therefore ignore the last column of $Y$ and the last row of
$ {\bf y}_0^{-1}$ when calculating the LHS, and contract $Y$ to the 
non-square $m$ by $m'$ matrix with elements given by the first of
the equations (\ref{defY1}), ${\bf y}_0^{-1}$ to the $m'$ by $m$ matrix
with elements given by the first of the equations (\ref{Y10}).


We  define the $m$ by $m'$ matrix  $A_+$ and the $m'$ by $m$ 
matrix $A_-^{}$ by
\be \label{defA}
(A_+)_{i,j} \eq  \frac{ 2  \,  \sin \theta_i }
{L( \cos \theta_i - \cos \theta'_j )}   \sep  
(A_-)_{i,j} \eq  \frac{ 2  \,  \sin \theta'_i }
	{L( \cos \theta'_i - \cos \theta_j )} \period  \ee
Then from (\ref{defY1}), (\ref{Y10}),
\be {\bf y}_1 = -i \tilde{V}_+(\alpha) A_+  \tilde{V}_-^{}(\beta) \sep
 {\bf y}_0^{-1} = -\i \tilde{U}_-(\beta)^{-1} A_-^{} \tilde{U}_+(\alpha)^{-1}
\comma  \ee
where we now take the $\tilde{U}_r, \tilde{V}_r, \tilde{X}_r$ to be the 
diagonal matrices defined by (\ref{uvt}), (\ref{defX}), but truncated
to the first $m(r)$ terms. 

 Hence 
 \bd  \tilde{\cal D} \eq
  - \tilde{U}_-(\beta)^{-1} {\cal D} \tilde{U}_-^{}(\beta) \comma 
  \ed
 where 
 \be \label{defcald}
{\cal D} \eq -  A_{-}^{}  X_+ (\alpha) A_+  X_{-}^{}(\beta) \period  \ee

{From} (\ref{detG2}), we obtain
\be \label {resdetG}
\det G  \eq \det\left(  \tilde{U}_+ (\alpha) \, F_+  \tilde{U}_{-}^{}(\beta) \right) 
\left[ \det ( I_{m'} + x {\cal D} ) \right] ^2 \comma \ee
where $I_{m'}$ is the identity $m'$ by $m'$ matrix.

The relations (\ref{resW}), (\ref{resW2}) therefore become
\be  \label{rescase1}
\widetilde{W} (\alpha, \beta,x) \eq \tilde{Z}_+(\alpha) \tilde{Z}_-^{}(\beta)
\det [ I_{m'} - x  A_{-}^{} X_+ (\alpha) A_+ X_{-}^{}(\beta) ] \period 
\ee
We have eliminated the square on the LHS  and used only matrices
of dimensions $m$ or $m'$, involving only $\theta_1, \ldots, \theta_m$, 
 $\theta'_1, \ldots, \theta'_{m'}$.

 \subsubsection*{The case $r=+$, $L$ odd}
 

Now (\ref{mmp}) gives $m = m' = (L-1)/2$, so the matrix
 $G$ has one exceptional row at $i = m+1$ and one exceptional column
 at $j=L$. The steps 
 corresponding to (\ref{6steps}) are
  \ba  
 \label{6steps2}
 1)  \; \; G_{ij} \rightarrow  &  G_{ij} + G_{L+1-i,j}  ,   & 
 1\leq i \leq m, \; 1 \leq j \leq L \nonumber \\
 2)  \; \; G_{ij} \rightarrow  &  G_{ij} -2\,  u_i G_{m+1,j} /u_{m+1} ,   & 
 1\leq i \leq m, \; 1 \leq j \leq L \nonumber \\
  3)  \; \; G_{ij}   \rightarrow  & G_{ij} + G_{i,L-j}  ,  &  m < j < L  
     \nonumber \\
4)  \; \;  G_{i,j}   \rightarrow  \! \!   & G_{i, j} - 2 \, u'_j G_{i,L}/u'_{L}  , 
  &   m < j < L   \nonumber \\
 5)  \; \; G_{i,j}   \rightarrow  & G_{i, j} -   u'_j G_{i,L}/u'_{L}  ,  
 &  1 \leq j \leq m   \\
  \nonumber \ea
The first two transformations correspond to pre-multiplying $G_r$  by 
some matrix, all the others 
to post-multiplying it. Steps (3) to (6) are for  all values of $1 \leq i \leq L$

Using (\ref{refF}), we find that the transformed $G$ has the structure
\be \label{valG2}
G = \left( \begin{array}
{ccc} Y  & 0  & 0\\ 
 \cdot \cdot  & 0  &  - u_{m+1} u'_L \\ {\cal T} & {\cal Z} &
 \cdot  \cdot 
 \end{array} \right) \comma \ee
 where $Y, {\cal T}, {\cal Z}$ are all  $m$ by $m'= m$ blocks and 
the middle row and last column have width one.
Hence \bd \det G \eq u_{m+1} u'_L  \, (\det Y) \, (\det {\cal Z})
\period \ed
we also find
\be \label{Y2nd}
Y_{ij} \eq \frac{- 4 \i \cos^2 (\theta_i/2) [ u_i u'_j \tan (\theta'_j /2)
+ x v_i v'_j \tan (\theta_i/2) ]}
{L( \cos \theta_i - \cos \theta'_j)}
\comma \ee
and again if we rearrange $\cal Z$ so that
$ {\cal Z}_{ij} = G_{L+1-i,L-j}$ for $1 \leq i,j \leq m$, then
\be {\cal Z} = -D^{-1} Y D' \comma \ee
with $D, D'$ given by 
(\ref{gdD}) for all $1 \leq j \leq m$. Further,
\be \label{Y102nd}({\bf y}_0^{-1})_{ij} \eq \frac {- 2 \i \sin \theta'_i }
{L (\cos \theta'_i - \cos \theta_j) u'_i u_j } \period \ee

{From} (\ref{defY1})  and (\ref{Y2nd}),
 ${\bf y}_0$ is different for the two cases ($L$ even and $L$ odd).
However, the wanted elements of ${\bf y}_1$ and of ${\bf y}_0^{-1}$ 
are the same,  so again  $\widetilde{W} (\alpha, \beta,x) $ is given by 
(\ref{rescase1}), with $A_+, A_-^{}$ given by (\ref{defA}).

{From} (\ref{uvt}), (\ref{defX}), the diagonal 
matrices $X_{\pm}$ have diagonal elements
 \be
[X_+(\alpha)]_{jj} = - \, 
  \frac{k'  \, \sin \theta_j 
\sinh( 2\alpha \lambda_j)}{\lambda_j \cosh( 2\alpha \lambda_j)+ 
(1-k' \cos \theta_j ) \sinh(2 \alpha \lambda_j )} \ee
 \bd
[X_-^{}(\beta)]_{jj} = - \, 
  \frac{k'  \, \sin \theta'_j 
\sinh( 2\beta \lambda'_j) }{\lambda'_j \cosh( 2\beta \lambda'_j)+ 
(1-k' \cos \theta'_j ) \sinh(2 \beta \lambda'_j )} \comma \ed
$\lambda_j$ being defined by (\ref{deflam}) and 
$\lambda'_j$ by the same equation with  $\theta_j$ 
replaced by $\theta'_j$.


 
\section{$\widetilde{W}$ in terms of orthogonal matrices $B_+,  B_-^{}$}
 

\setcounter{equation}{0}

In this section we take the result (\ref{rescase1})
and write it in terms, not of the matrices $A_+,  A_-^{}$, but of related
orthogonal matrices $B_+, B_-^{}$.

In both the $L$ even and odd cases, the matrices ${\bf y}_0^{-1}, 
{\bf y}_0^T$ above
have a very similar structure, differing only by diagonal equivalence
transformations. We can therefore construct orthogonal 
matrices $B_+, B_-^{}$ that differ from $A_+, A_-^{}$
only by such diagonal equivalence transformations.

Let $E, E'$ be diagonal matrices with diagonal elements
\ba E_{jj}  &  \! \! \! \! \! \! \! = \sin \theta_j \sep E'_{jj} = \sin \theta'_j 
\; \; & {\rm if} \; \;  L = \; {\rm even} \, , \nonumber \\
 E_{jj}  & = \tan ( \theta_j/2 )\sep E'_{jj} = \tan (\theta'_j/2 )
\; \; & {\rm if} \; \;  L = \; {\rm odd} \, . \ea

Then whether $L$ is even or odd, we find from
(\ref{defY1}), (\ref{gdD}), (\ref{Y10}), (\ref{defA}), (\ref{Y2nd}),
(\ref{Y102nd}) that
\be {\bf y}_0^{-1} = \i \, D' A_+^T D^{-1} \sep 
{\bf y}_0 = -\i \, E^{-1} A_+ E' \sep
A_-^T = -D^{-1} A_+ D' \comma \ee
${\bf y}_0^{-1}$ being the left-inverse of ${\bf y}_0$.

Hence we can define matrices $B_{+}, B_{-}^{}$:
\be B_{+} = (DE)^{-1/2} A_+ (D'E')^{1/2} \sep 
B_{-}^{} = (E'/D')^{1/2} A_-^{} (D/E)^{1/2} \ee
such that
\be \label{propB}
B_{-}^{} B_{- }^{\, T} =  B_{+}^{\, T} B_{+} = I_{m'} \sep B_{-}^{} \, = 
- B_{+}^{ \, T} \comma \ee
so, even though they are not necessarily square, they are orthogonal
in this sense. Their elements are
\ba (B_+)_{ij} = &  - (B_-^{})_{ji} = & \frac{ 2 \, \sin \theta'_j}
{L (\cos \theta_i - \cos \theta'_j)}
\; \; \; \; {\rm if} \; \; L = \; {\rm even} \, , \nonumber \\ 
 (B_+)_{ij} = & - (B_-^{})_{ji} = & \frac{ 4 \, \cos(\theta_i/2) 
\sin ( \theta'_j/2)}
{L (\cos \theta_i - \cos \theta'_j)}
\; \; {\rm if} \; \; L = \; {\rm odd} \, . 
\ea

Set
\be E_+ = E \sep E_-^{} = E'^{-1} \period \ee
then we can write the result (\ref{rescase1}) as
\be  \label{finres}
\widetilde{W} (\alpha, \beta,x) \eq \tilde{Z}_+(\alpha) \tilde{Z}_-^{}(\beta)
\det [ I_{m'} - x  X_+ (\alpha) \,   E_+ B_+  X_{-}^{}(\beta)   \, E_-^{} 
  B_{-}^{}] \period 
\ee

We can slightly generalize this result.
Write  $\widetilde{W}(\alpha,\beta,x)$ as  
$\widetilde{W}_+(\alpha,\beta,x)$, and define a new function
$\widetilde{W}_-^{}(\alpha,\beta,x)$ to be given by (\ref{defW}) with 
$v_+, v_-^{}$ interchanged. Thus (\ref{defZ}), (\ref{defW})
become
\be \label{defZW2}
\tilde{Z}_r(\alpha) = v_r^{\dagger} \e^{\alpha {\cal H}} v_r
\sep \widetilde{W}_r(\alpha, \beta ,x) = v_r^{\dagger}  \e^{\alpha {\cal H}}
 \e^{\rho {\cal J}} S_1  \e^{\beta {\cal H}} v_{-r} \period \ee
Then we can generalize (\ref{finres}) to
\be  \label{finres2}
\widetilde{W}_r (\alpha, \beta,x) \eq \tilde{Z}_r(\alpha) \tilde{Z}_{-r}^{}(\beta)
\det [ I - x X_r (\alpha) \,  \, E_r \,  B_r 
X_{-r}^{}(\beta)  E_{-r}^{}   B_{-r}^{}  ] \comma 
\ee
for $r = \pm$, where $I$ is the identity matrix of dimension $m(r)$. For $r=+$,
(\ref{finres2}) is (\ref{finres}). For $r=-$ it can be deduced from it
by taking the hermitian conjugate of the definition
(\ref{defZW2}) of  $\widetilde{W}_-^{} $  and interchanging $\alpha$ with 
$\beta$.  We also need the general identity
 \bd \det (I+AB) = \det  (I + BA) \comma \ed
 which is true for all matrices $A,B$ even when they are non-square,


 

 \section{Conclusions}
 
 
 We have considered the zero-field Ising model on a cylindrical
  lattice with fixed spin boundary condtions on the top and bottom rows, 
  and have replaced the transfer matrix product by the exponential of the 
  associated  hamiltonian. This leaves the eigenvectors and spontaneous 
  magnetization  unchanged.
  
We have then used the Clifford algebra technique of 
Kaufman\cite{Kaufman1949}  to evaluate as $L$ by $L$ determinants
the partition function $\tilde{Z}$, and the partition function
$\widetilde{W}$ with a single-spin operator
$S_1$ included.  In section 5 we show how  to use Szeg{\H o}'s theorem 
on  Toeplitz matrices to then obtain the spontaneous  magnetization.

Much of this merely parallels Yang's derivation\cite{Yang1952}, but in 
section 6 we go on to reduce  $\tilde{Z}$ and 
$\widetilde{W}$ to determinants of size $m$ or $m'$ where
$L-2 \leq 2m \leq L$. 

In this form we can compare the Ising model results
for $\tilde{Z}$ with those of the  $N$-state superintegrable chiral Potts 
model, taking the hamiltonian limit 
therein.\cite{Baxter1989,Baxter2008b} The Ising model is the 
 superintegrable chiral Potts model with $N=2$, and the results
 of course agree.

All this leads up to the question whether we 
can generalize the Ising result (\ref{rescase1}) or (\ref{finres2})
 for $\widetilde{W}$ 
to the $N$-state superintegrable chiral Potts model, for general $N$.
This should open up the possibility 
of an algebraic derivation of the spontaneous magnetization of that model.
We have numerical evidence that strongly suggests the answer to the 
question is yes, and that one can make a fairly immediate generalization 
of the form of the result given in (\ref{finres2}) in terms of the orthogonal
matrices $B_+, B_-^{}$. We shall present the conjecture in an 
accompanying paper.\cite{Baxter2008b}



\begin{thebibliography}{9999}
 
  \bibitem{Onsager1944} Onsager,~L.: Crystal statistics. I. A 
  two-dimensional model with an order-disorder transition.
   { Phys. Rev} {\bf 65}, 117--149 (1944)
  
   \bibitem{Kaufman1949} Kaufman,~B.: Crystal statistics. II. Partition
   function evaluated by spinor analysis.
    { Phys. Rev} {\bf 76}, 1232--1243 (1949)
   
   
  \bibitem{Onsager1949} Onsager, ~L. In: Proceedings of the IUPAP
conference on statistical mechanics, ``Discussione e observazioni'',
 {Nuovo Cimento (Suppl), Series 9},
{\bf 6}, 261 (1949)

 
  \bibitem{MPW1963} Montroll,~E.~W., Potts,~R.~B.,   Ward,~J.~C.: 
  Correlations and spontaneous magnetization of the two-dimensional Ising 
  model.  { J. Math. Phys.} {\bf 4}, 308--322 (1963)
  
 \bibitem{Yang1952} Yang,~C.~N.:  The spontaneous magnetization of a
two-dimensional Ising model.  { Phys. Rev.} {\bf 85}, 808--816 (1952)


 \bibitem{Baxter2005a} Baxter,~R.~J.:  Derivation of the order parameter 
 of the chiral Potts model. {Phys. Rev. Lett.} {\bf 94}, 130602 
(2005)

  \bibitem{Baxter2005b} Baxter,~R.~J.:  The order parameter of the 
 chiral Potts model. {J. Stat. Phys.} {\bf 120}, 1--36 (2005)
 
  
 
\bibitem{Szego1958} Grenander,~U.,  Szeg{\H o},~G.: Toeplitz forms and
 their applications. Univ. Calif. Press, Berkeley, (1958)

\bibitem{Baxter1982} Baxter,~R.~J.: Exactly solved models in statistical 
mechanics.  Academic Press,  London/San Diego1989; Dover Publications, 
Mineola, NY (2007)


  \bibitem{Baxter1977}  Baxter,~R.~J.:  Corner transfer matrices of the
   eight-vertex model. II. The Ising model case.
 { J. Stat. Phys.} {\bf 17}, 1--14 (1977)


 \bibitem{Baxter1989} Baxter,~R.~J.:
 Superintegrable chiral Potts model: thermodynamic properties, 
 an ``inverse'' model, and a simple associated hamiltonian. 
 {J. Stat. Phys.}  {\bf 57}, 1--39 (1989)

\bibitem{Baxter2008b}  Baxter,~R.~J.:  A conjecture for the
superintegrable chiral Potts model.   {J. Stat. Phys.}  {\bf 132}, ??--?? 
(2008)


 \end{thebibliography}
 \end{document}